\documentclass[english,aps,manuscript]{revtex4}
\usepackage[T1]{fontenc}
\usepackage[latin9]{inputenc}
\usepackage[active]{srcltx}
\usepackage{amsmath}
\usepackage{esint}

\makeatletter

\DeclareRobustCommand{\greektext}{%
  \fontencoding{LGR}\selectfont\def\encodingdefault{LGR}}
\DeclareRobustCommand{\textgreek}[1]{\leavevmode{\greektext #1}}
\DeclareFontEncoding{LGR}{}{}
\DeclareTextSymbol{\~}{LGR}{126}
\newcommand{\lyxmathsym}[1]{\ifmmode\begingroup\def\b@ld{bold}
  \text{\ifx\math@version\b@ld\bfseries\fi#1}\endgroup\else#1\fi}

\@ifundefined{textcolor}{}
{%
 \definecolor{BLACK}{gray}{0}
 \definecolor{WHITE}{gray}{1}
 \definecolor{RED}{rgb}{1,0,0}
 \definecolor{GREEN}{rgb}{0,1,0}
 \definecolor{BLUE}{rgb}{0,0,1}
 \definecolor{CYAN}{cmyk}{1,0,0,0}
 \definecolor{MAGENTA}{cmyk}{0,1,0,0}
 \definecolor{YELLOW}{cmyk}{0,0,1,0}
 }

\makeatother

\usepackage{babel}
\begin{document}

\title{Geometric Thermodynamics of Kerr-AdS black hole with a Cosmological
Constant as State Variable}

\author{Alexis Larrañaga}

\address{National Astronomical Observatory. National University of Colombia.}

\author{Sindy Mojica}

\address{Physics Department. National University of Colombia.}
\begin{abstract}
The thermodynamics of the Kerr-AdS black hole is reformulated within
the context of the formalism of geometrothermodynamics (GTD). Different
choices of the metric in the equilibrium states manifold are used
in order to study the phase transitions as a divergence of the thermodynamical
curvature scalar.


\end{abstract}
\maketitle

\section{Introduction}

The thermodynamics of black holes has been studied extensively since
the work of Hawking \cite{key-1}. The notion of critical behaviour
for black holes has arisen in several contexts from the Hawking-Page
\cite{key-2} phase transition in anti-de-Sitter background to the
pioneering work by Davies \cite{key-3} on the thermodynamics of Kerr-Newman
black holes and the idea of the extremal limit of various black hole
families regarded as genuine critical points \cite{key-4,key-5,key-6}.
Recently, some authors have considered the cosmological constant $\Lambda$
as a dynamical variable \cite{lambda1,lambda2} and it has further
been suggested that it is better to consider $\Lambda$ as a thermodynamic
variable, \cite{lambda3,lambda4,lambda5,lambda6,quevedo08-1}. Physically,
$\Lambda$ is interpreted as a thermodynamic pressure in \cite{entalpia1,entalpia2},
fact that is consistent with the observation in \cite{entalpia3,dolan,cvetic}
that its conjugate thermodynamic variable is proportional to a volume. 

The use of geometry in statistical mechanics was pioneered by Ruppeiner
\cite{rup79} and Weinhold \cite{wei1}, who suggested that the curvature
of a metric defined on the space of parameters of a statistical mechanical
theory could provide information about its phase structure. When this
treatment is applied to the study of black hole thermodynamics, some
puzzling anomalies appear. A possible solution was suggested by Quevedo\textquoteright{}s
geometrothermodynamics (GTD) whose starting point \cite{quev07} was
the observation that standard thermodynamics is invariant with respect
to Legendre transformations. The formalism of GTD indicates that phase
transitions occur at those points where the thermodynamic curvature
scalar is singular.

In this paper we apply the GTD formalism to the Kerr-AdS black hole
to investigate the behavior of the thermodynamical curvature. As is
well known, a black hole with a positive cosmological constant has
both a cosmological horizon and an event horizon. These two surfaces
have, in general, different Hawking temperatures, which complicates
any thermodynamical treatment. Therefore, we will focus on the case
of a negative cosmological constant, though many of the conclusions
are applicable to the positive $\Lambda$ case. Evenmore, the negative
$\Lambda$ case is of interest for studies on AdS/CFT correspondence
and the considerations of this work are likely to be relevant in those
studies.

\section{Geometrothermodynamics in Brief}

The formulation of GTD is based on the use of contact geometry as
a framework for thermodynamics. The $(2n+1)$-dimensional thermodynamic
phase space $\mathcal{T}$ is coordinatized by the thermodynamic potential
$\Phi$, the extensive variables $E^{a}$, and the intensive variables
$I^{a}$, with $a=1,...,n$. We define on $\mathcal{T}$ a non-degenerate
metric $G=G(Z^{A})$ with $Z^{A}=\{\Phi,E^{a},I^{a}\}$, and the Gibbs
1-form $\Theta=d\Phi-\delta_{ab}I^{a}dE^{b}$ with $\delta_{ab}={\rm diag}(1,1,...,1)$.
If the condition $\Theta\wedge(d\Theta)^{n}\neq0$ is satisfied, the
set $(\mathcal{T},\Theta,G)$ defines a contact Riemannian manifold.
The Gibbs 1-form is invariant with respect to Legendre transformations,
while the metric $G$ is Legendre invariant if its functional dependence
on $Z^{A}$ does not change under a Legendre transformation. This
invariance guarantees that the geometric properties of $G$ do not
depend on the thermodynamic potential used in its construction. 

Now, we define the $n$-dimensional subspace of equilibrium thermodynamic
states, $\mathcal{E}\subset\mathcal{T}$, by means of the smooth mapping
\begin{eqnarray}
\varphi:\ \mathcal{E} & \longrightarrow & \mathcal{T}\nonumber \\
(E^{a}) & \longmapsto & \left(\Phi,E^{a},I^{a}\right)
\end{eqnarray}
with $\Phi=\Phi(E^{a})$, and the condition $\varphi^{*}(\Theta)=0$,
which gives the first law of thermodynamics 
\begin{equation}
d\Phi=\delta_{ab}I^{a}dE^{b}
\end{equation}

and the conditions for thermodynamic equilibrium (the intensive thermodynamic
variables are dual to the extensive ones), 
\begin{equation}
\frac{\partial\Phi}{\partial E^{a}}=\delta_{ab}I^{b}.
\end{equation}
The mapping $\varphi$ defined above implies that we know the equation
$\Phi=\Phi(E^{a})$ explicitly. It is known as the fundamental equation,
and from it can be derived all the equations of state. The second
law of thermodynamics is equivalent to the convexity condition on
the thermodynamic potential, 
\begin{equation}
\partial^{2}\Phi/\partial E^{a}\partial E^{b}\geq0.
\end{equation}

Since the thermodynamic potential satisfies the homogeneity condition
$\Phi(\lambda E^{a})=\lambda^{\beta}\Phi(E^{a})$ for constant parameters
$\lambda$ and $\beta$, it satisfies Euler's identity, 
\begin{equation}
\beta\Phi(E^{a})=\delta_{ab}I^{b}E^{a},
\end{equation}
and using the first law of thermodynamics, this gives the Gibbs-Duhem
relation, 
\begin{equation}
(1-\beta)\delta_{ab}I^{a}dE^{b}+\delta_{ab}E^{a}dI^{b}=0.
\end{equation}

Defining a non-degenerate metric structure $g$ on ${\cal E}$ that
is compatible with a metric $G$ on ${\cal T}$, we state that a thermodynamic
system is described by the thermodynamical metric $G$ \cite{quev07}
if it is invariant with respect to transformations which do not modify
the contact structure of ${\cal T}$ . In particular, $G$ must be
invariant with respect to Legendre transformations in order for GTD
to be able to describe thermodynamic properties in terms of geometric
concepts independent of the the thermodynamic potential used. A partial
Legendre transformation is written as

\begin{equation}
Z^{A}\rightarrow\tilde{Z}^{A}=\left\{ \tilde{\Phi},\tilde{E}^{a},\tilde{I}^{a}\right\} 
\end{equation}
where

\begin{equation}
\begin{cases}
\Phi & =\tilde{\Phi}-\delta_{kl}\tilde{E}^{k}\tilde{I}^{l}\\
E^{i} & =-\tilde{I}^{i}\\
E^{j} & =\tilde{E}^{i}\\
I^{i} & =\tilde{E}^{i}\\
I^{j} & =\tilde{I}^{j},
\end{cases}
\end{equation}

with $i\cup j$ any disjoint decomposition of the set of indices $\left\{ 1,2,...,n\right\} $
and $k,l=1,...,i$. As is shown in \cite{quev07}, a Legendre invariant
metric $G$ induces a Legendre invariant metric $g$ on ${\cal E}$
defined by the pullback $\varphi^{*}$ as $g=\varphi^{*}(G)$. There
is a vast number of metrics on ${\cal T}$ that satisfy the Legendre
invariance condition. The results of Quevedo et al. \cite{quevedo08,quevedo10,quevedo09}
show that phase transitions occur at those points where the thermodynamic
curvature is singular and that the metric structure of the phase manifold
${\cal T}$ determines the type of systems that can be described by
a specific thermodynamic metric. For instance, a pseudo-Euclidean
structure 
\begin{equation}
G=\Theta^{2}+(\delta_{ab}E^{a}I^{b})(\eta_{cd}dE^{c}dI^{d})\label{eq:Gmetric}
\end{equation}
 with $\eta_{cd}=\mbox{diag}\left(-1,1,1,...,1\right)$is Legendre
invariant because of the invariance of the Gibbs 1-form and induces
on ${\cal E}$ the Quevedo's metric 
\begin{equation}
g=\left(E^{f}\frac{\partial\Phi}{\partial E^{f}}\right)\left(\eta_{ab}\delta^{bc}\frac{\partial^{2}\Phi}{\partial E^{c}\partial E^{d}}dE^{a}dE^{d}\right)\label{eq:metricg}
\end{equation}

which describes systems characterized with second order phase transitions.
On the other hand, an euclidean structure 

\begin{equation}
G=\Theta^{2}+(\delta_{ab}E^{a}I^{b})(\delta_{cd}dE^{c}dI^{d})\label{eq:Gmetric1}
\end{equation}

it is also Legendre invariant and induces on ${\cal E}$ the metric

\begin{equation}
g=\left(E^{f}\frac{\partial\Phi}{\partial E^{f}}\right)\left(\frac{\partial^{2}\Phi}{\partial E^{c}\partial E^{d}}dE^{c}dE^{d}\right)\label{eq:metricg1}
\end{equation}

which describes systems with first order phase transitions.

\section{The Kerr-AdS Black Hole}

The Einstein action with cosmological constant $\Lambda$ term is
given by

\begin{equation}
\mathcal{A}=\frac{1}{16\pi}\int d^{4}x\sqrt{-g}\left(R-2\Lambda\right),
\end{equation}
and the general solution representing a black hole is given by the
Kerr-AdS solution 

\begin{eqnarray}
ds^{2} & = & -\frac{\Delta_{r}}{\rho^{2}}\left(dt-\frac{a\sin^{2}\theta}{\Xi}d\varphi\right)^{2}+\frac{\Delta_{\theta}\sin^{2}\theta}{\rho^{2}}\left(adt-\frac{r^{2}+a^{2}}{\Xi}d\varphi\right)^{2}\nonumber \\
 &  & +\rho^{2}\left(\frac{dr^{2}}{\Delta_{r}}+\frac{d\theta^{2}}{\Delta_{\theta}}\right)
\end{eqnarray}
where 

\begin{equation}
\Delta_{r}=\left(r^{2}+a^{2}\right)\left(1-\frac{\Lambda r^{2}}{3}\right)-2mr
\end{equation}

\begin{equation}
\Delta_{\theta}=1+\frac{\Lambda a^{2}}{3}\cos^{2}\theta
\end{equation}

\begin{equation}
\rho^{2}=r^{2}+a^{2}\cos^{2}\theta
\end{equation}

and

\begin{equation}
\Xi=1+\frac{\Lambda a^{2}}{3}.
\end{equation}

The physical parameters of the black hole can be obtained by means
of Komar integrals using the Killing vectors $\frac{\partial_{t}}{\Xi}$
and $\partial_{\varphi}$. In this way, one obtains the mass of the
black hole

\begin{equation}
M=\frac{m}{\Xi^{2}}
\end{equation}
and its angular momentum 
\begin{equation}
J=aM=a\frac{m}{\Xi^{2}}.
\end{equation}

The horizons are given by the roots of 

\begin{equation}
\Delta_{r}=0.\label{eq:horizon}
\end{equation}

In particular, the largest positive root located at $r=r_{+}$ defines
the event horizon with an area

\begin{eqnarray}
A & = & 4\pi\frac{\left(r_{+}^{2}+a^{2}\right)}{\Xi}.\label{eq:area}
\end{eqnarray}

The Smarr formula for the Kerr-AdS black hole gives the relation

\begin{equation}
M^{2}=J^{2}\left(\frac{\pi}{S}-\frac{\Lambda}{3}\right)+\frac{S^{3}}{4\pi^{3}}\left(\frac{\pi}{S}-\frac{\Lambda}{3}\right)^{2}\label{eq:fundamentalM}
\end{equation}
that corresponds to the fundamental thermodynamical equation $M=M\left(S,J,\Lambda\right)$
which relates the total mass $M$ of the black hole with the extensive
variables, entropy $S=\frac{A}{4}$, angular momentum $J$ and cosmological
constant$\Lambda$, and from which all the thermodynamical information
can be derived. 

In the geometric formulation of thermodynamics we will choose the
extensive variables as $E^{a}=\left\{ S,J,\Lambda\right\} $ and the
corresponding intensive variables as $I^{a}=\left\{ T,\Omega,\Psi\right\} $,
where $T$ is the temperature, $\Omega$ is the angular velocity and
$\Psi$ is the generalized variable conjugate to the state parameter
$\Lambda$. Therefore, the coordinates that we will use in the 7-dimensional
thermodynamical space ${\cal T}$ are $Z^{A}=\left\{ M,S,J,\Lambda,T,\Omega,\Psi\right\} $.
The contact structure of ${\cal T}$ is generated by the 1-form

\begin{equation}
\Theta=dM-TdS-\Omega dJ-\Psi d\Lambda.
\end{equation}

To obtain the induced metric in the space of equilibrium states ${\cal E}$
we will introduce the smooth mapping

\begin{equation}
\varphi:\{S,J,\Lambda\}\longmapsto\left\{ M(S,J,\Lambda),S,J,\Lambda,T\left(S,J,\Lambda\right),\Omega\left(S,J,\Lambda\right),\Psi\left(S,J,\Lambda\right)\right\} 
\end{equation}

along with the condition $\varphi^{*}(\Theta)=0$, that corresponds
to the first law $dM=TdS+\Omega dJ+\Psi d\Lambda$. This condition
also gives the relation between the different variables with the use
of the fundamental relation (\ref{eq:fundamentalM}). The Hawking
temperature is evaluated as

\begin{equation}
T=\frac{\partial M}{\partial S}=\frac{S^{2}}{8\pi^{3}M}\left(\frac{\pi}{S}-\frac{\Lambda}{3}\right)\left(\frac{\pi}{S}-\Lambda\right)-\frac{\pi J^{2}}{2MS^{2}},
\end{equation}

the angular velocity is

\begin{equation}
\Omega=\frac{\partial M}{\partial J}=\frac{J}{M}\left(\frac{\pi}{S}-\frac{\Lambda}{3}\right)
\end{equation}

and the conjugate variable to $\Lambda$ is 

\begin{equation}
\Psi=\frac{\partial M}{\partial\Lambda}=-\frac{S^{3}}{12\pi^{3}M}\left(\frac{\pi}{S}-\frac{\Lambda}{3}\right)-\frac{J^{2}}{6M}.
\end{equation}

As can be seen, $\Psi$ has dimensions of a volume. In fact, in the
limit of non-rotating black hole, $J\rightarrow0$, we have $\Psi=-\frac{4}{3}r_{+}^{3}$
(see \cite{LarCar}) and it can be interpreted as an effective volume
excluded by the horizon, or alternatively a regularised version of
the diff{}erence in the total volume of space with and without the
black hole present \cite{entalpia1,entalpia2,entalpia3}. Since the
cosmological constant $\lyxmathsym{\textgreek{L}}$ behaves like a
pressure and its conjugate variable as a volume, the term $\Psi d\Lambda$
has the correct dimensions of energy and is the analogue of $VdP$
in the first law. This suggests that after expanding the set of thermodynamic
variables to include the cosmological constant, the mass $M$ of the
AdS black hole should be interpreted as the \emph{enthalpy} rather
than as the total energy of the spacetime. 

${\cal T}$ becomes a Riemannian manifold by defining the metric (\ref{eq:Gmetric}),

\begin{equation}
G=\left(dM-TdS-\Omega dJ-\Psi d\Lambda\right)^{2}+\left(ST+\Omega J+\Psi\Lambda\right)\left(-dSdT+dJd\Omega+d\Lambda d\Psi\right).
\end{equation}

$G$ has non-zero curvature and its determinant is $\det\left[G\right]=-\frac{\left(ST+\Omega J+\Psi\Lambda\right)^{6}}{64}$.
Equation (\ref{eq:metricg}) let us define the induced metric structure
on ${\cal E}$ as

\begin{equation}
g=\left(SM_{S}+JM_{J}+\Lambda M_{\Lambda}\right)\left(\begin{array}{ccc}
-M_{SS} & 0 & 0\\
0 & M_{JJ} & M_{J\Lambda}\\
0 & M_{J\Lambda} & M_{\Lambda\Lambda}
\end{array}\right),\label{eq:g1}
\end{equation}
where subscripts represent partial derivative with respect to the
corresponding coordinate. Note that the determinant of this metric
is 
\begin{equation}
\det\left[g\right]=M_{SS}\left(M_{J\Lambda}^{2}-M_{JJ}M_{\Lambda\Lambda}\right)\left(SM_{S}+JM_{J}+\Lambda M_{\Lambda}\right)^{3}.\label{eq:detg}
\end{equation}

We can also define an euclidean metric (\ref{eq:Gmetric1}) on ${\cal T}$,
but there are no phase transitions associated with this metric.

\section{Phase Transitions and The Curvature Scalar}

Phase transitions are an interesting subject in the study of black
holes thermodynamics since there is no unanimity in their definition.
In ordinary thermodynamics, phase transitions are defined by looking
for singular points in the behavior of thermodynamical variables.
Davis \cite{key-3,davisthermo} shows that the diveregences in the
heat capacity indicate phase transitions. For example, using equation
(\ref{eq:fundamentalM}) we have that the heat capacity for the Kerr-AdS
black hole is

\begin{equation}
C=T\frac{\partial S}{\partial T}=\frac{M_{S}}{M_{SS}}
\end{equation}

\begin{equation}
C=\frac{S\left(\frac{\pi}{S}-\frac{\Lambda}{3}\right)\left(\frac{\pi}{S}-\Lambda\right)-\frac{4\pi^{4}J^{2}}{S^{3}}}{\left(\frac{\pi}{S}-\frac{\Lambda}{3}\right)\left(\frac{\pi}{S}-2\Lambda\right)-\frac{\pi}{S}\left(\frac{\pi}{S}-\Lambda\right)+\frac{8\pi^{3}}{S^{2}}\left(\frac{\pi J^{2}}{S^{2}}-ST^{2}\right)}.
\end{equation}
Thus, one can expect that phase transitions occur at the divergences
of $C$, i.e. at points where $M_{SS}=0$. For negative $\Lambda$
the divergence of $C$ corresponds to the generalization of the well
known Hawking-Page transition \cite{key-2}. In GTD the apparition
of phase transitions appears to be related with the divergences of
the curvature scalar $R$ in the space of equilibrium states ${\cal E}$.
To understand this relation, remember that $R$ always contains the
determinant of the metric $g$ in the denominator and, therefore,
the zeros of $\det\left[g\right]$ could lead to curvature singularities
(if those zeros are not canceled by the zeros of the numerator). 

Here we have considered the metric $g$ given in (\ref{eq:g1}) and
its determinant is proportional to $M_{SS}$ as shown in equation
(\ref{eq:detg}), making clear the coincidence with the divergence
of the heat capacity and the apparition of a second order phase transition
that corresponds to the generalization of the Hawking-Page result.
There is also a factor of $\left(M_{J\Lambda}^{2}-M_{JJ}M_{\Lambda\Lambda}\right)$
in the determinant which codifies the information of non-constant
$\Lambda$. Note that the interesting second derivatives of the thermodynamic
potential are

\[
M_{SS}=\frac{144\pi^{7}J^{4}\left(9\pi-4\Lambda S\right)+24\pi^{3}J^{2}S^{2}\left(3\pi-2\Lambda S\right)\left(\Lambda S-3\pi\right)^{2}+S^{4}\left(\Lambda S-3\pi\right)^{3}\left(\Lambda S+\pi\right)}{8\pi^{3/2}S^{4}\left[\frac{\left(\Lambda S-3\pi\right)\left(S^{2}\left(\Lambda S-3\pi\right)-12\pi^{3}J^{2}\right)}{S}\right]^{3/2}}
\]

\[
M_{JJ}=-\frac{2\pi^{3/2}\left(\Lambda S-3\pi\right)^{3}}{\left[\frac{\left(\Lambda S-3\pi\right)\left(S^{2}\left(\Lambda S-3\pi\right)-12\pi^{3}J^{2}\right)}{S}\right]^{3/2}}
\]

\[
M_{\Lambda\Lambda}=-\frac{6\pi^{9/2}J^{4}}{\left[\frac{\left(\Lambda S-3\pi\right)\left(S^{2}\left(\Lambda S-3\pi\right)-12\pi^{3}J^{2}\right)}{S}\right]^{3/2}}
\]

\[
M_{J\Lambda}=\frac{12\pi^{9/2}J^{3}\left(\Lambda S-3\pi\right)}{S\left[\frac{\left(\Lambda S-3\pi\right)\left(S^{2}\left(\Lambda S-3\pi\right)-12\pi^{3}J^{2}\right)}{S}\right]^{3/2}}.
\]

As it can be seen, for negative values of $\Lambda$ the factor $\left(M_{J\Lambda}^{2}-M_{JJ}M_{\Lambda\Lambda}\right)$
is always positive. Therefore, we conclude that considering $\Lambda$
as a new thermodynamical state parameter do not produce new phase
transitions in Kerr-AdS black hole.

\section{Conclusion}

Quevedo's geometrothermodynamics describes in an invariant manner
the properties of thermodynamic systems using geometric concepts.
It indicates that phase transitions would occur at those points where
the thermodynamic curvature $R$ is singular. Following Quevedo, the
choice of metric given in equation (\ref{eq:metricg}) apparently
describe second order phase transitions.

In this work we applied the GTD formalism to the Kerr-AdS black hole,
considering the cosmological constant as a new thermodynamical state
variable. In this aproach, the total mass of the black hole is interpreted
as the total enthalpy of the system. Then, we obtain a curvature scalar
that diverges exactly at the point where the Hawking-Page phase transition
occurs. Since we use a metric of the form given in (\ref{eq:metricg})
we conclude that this is a second order phase transition. It is also
important to note that the consideration of $\Lambda$ as a thermodynamical
variable does not include new phase transitions in the system.

It is clear that the phase manifold in the GTD formalism contains
information about thermodynamic systems; however, it is not clear
now where is encoded the thermodynamic information. A more detailed
investigation along these lines will be reported in the future.\\

\emph{Acknowledgements}

This work was supported by the Universidad Nacional de Colombia. Hermes
Project Code 13038.

\end{document}